\documentclass{PoS}
\usepackage[utf8]{inputenc}
\usepackage{fontenc,times,mathptmx}
\usepackage{bm}
\usepackage{amsmath,amssymb,mathrsfs}
\usepackage{enumerate}
\usepackage{graphicx}
\usepackage{multirow}
\usepackage[squaren,Gray]{SIunits}

\def\plots{0.6}

\newcommand{\bib}[6]{\bibitem{#1}#2, \emph{#3}, \emph{#4} \textbf{#5} #6}
\newcommand{\cf}{\textit{cf.}~}
\newcommand{\diff}{\mathrm{d}}
\newcommand{\Diff}{\mathrm{D}}
\newcommand{\equ}[1]{\begin{equation}#1\end{equation}}
\newcommand{\etal}{\textit{et al.}~}
\newcommand{\etc}{\textit{etc.}~}
\newcommand{\ev}{\electronvolt}
\newcommand{\ie}{\textit{i.e.}~}
\newcommand{\mQCD}{\mathrm{QCD}}
\newcommand{\mQED}{\mathrm{QED}}
\newcommand{\bigo}{\mathrm{O}}
\newcommand{\SU}{\mathrm{SU}}
\newcommand{\U}{\mathrm{U}}

\title{Electromagnetic corrections to light hadron masses}
\ShortTitle{Electromagnetic corrections to light hadron masses}
\author{\speaker{A. Portelli}$\,^{b}$, S. D\"urr$\,^{a,d}$, Z. Fodor$\,^{a,c,d}$, J. Frison$\,^{b}$,
C. H\"olbling$\,^{a}$, S.D. Katz$\,^{a,c}$, \mbox{S. Krieg}$\,^{a,d}$, T. Kurth$\,^{a}$, 
L. Lellouch$\,^{b}$, T. Lippert$\,^{a}$, 
 K.K. Szab\'o$\,^{a}$ and A. Ramos$\,^{b}$
\mbox{(Budapest-Marseille-Wuppertal Collaboration)}\\
$\,^{a}$Bergische Universität Wuppertal, Gau\ss str. 20, D-42097 Wuppertal, Germany\\
$\,^{b}$Centre de Physique Th\'eorique, Luminy, Case 907, F-13288 Marseille cedex 9, France
\thanks{CPT is research unit UMR 6207 of the CNRS and of the universities 
Aix-Marseille II, Aix-Marseille I and
Sud Toulon-Var, and is affiliated with the FRUMAM}\\
$\,^{c}$Institute for Theoretical Physics, E\"otv\"os University, H-1117 Budapest, Hungary\\
$\,^{d}$J\"ulich Supercomputing Centre, Forschungszentrum J\"ulich, D-52425 J\"ulich, Germany\\
~\\
E-mail : \email{portelli@cpt.univ-mrs.fr}}
\author

\abstract{At the precision reached in current lattice QCD calculations,
electromagnetic effects are becoming numerically relevant. We will
present preliminary results for electromagnetic corrections to light
hadron masses, based on simulations in which a $\U(1)$ degree of freedom
is superimposed on $N_f=2+1$ QCD configurations from the BMW collaboration.}

\FullConference{The XXVIII International Symposium on Lattice Field Theory\\
June 14-19,2010\\
Villasimius, Sardinia Italy}

\begin{document}

%%%%%%%%%%%%%%%%%%%%%%%%%%%%%%%%%%%%%%%%%%%%%%%%%%%%%%%%%%%%%%%%%%%%%%%%%
\section{Motivation}
Isospin is a near symmetry of the hadron spectrum because the strong interaction only
distinguishes quark flavors through their masses and the mass difference between up and
down quarks is small. This symmetry is broken by :

\begin{enumerate}[(i)]
\item the mass difference $m_{u}-m_{d}$ (\emph{mass isospin breaking})
\item the difference in the charge of the $u$ and the $d$ quark (\emph{electromagnetic isospin breaking})
\end{enumerate}
\begin{center}
\begin{tabular}{|l|c|c|}
		\cline{2-3}
		\multicolumn{1}{c|}{~} & $u$ & $d$\\
		\hline
		Mass ($\mega\ev$) \cite{PDG10}& $2.49^{+0.81}_{-0.79}$ & $5.05^{+0.75}_{-0.95}$\\
		Charge & $\frac{2}{3}e$ & $-\frac{1}{3}e$\\	
		\hline
\end{tabular}
\end{center}

\noindent These effects are expected to be at the percent level. The size of mass breaking is
the mass difference $m_{u}-m_{d}$ relatively to a typical QCD scale $\Lambda_{\mQCD}$ and the
order of electromagnetic breaking is the fine structure constant at zero momentum
$\alpha=\frac{e^{2}}{4\pi}\simeq\frac{1}{137}$.

These effects imply mass splittings inside isospin multiplets.
Although these effects are small, they have important consequences. For example, the fact that the
neutron is heavier than the proton guarantees the stability of matter. Another interesting isospin breaking quantity is the
\emph{absolute correction to Dashen's theorem} :
\equ{\Delta_{A}D = \Delta_{\mathrm{EM}}M_{K}^{2}-\Delta_{\mathrm{EM}}M_{\pi}^{2}}
where :
\equ{\Delta_{\mathrm{EM}}M_{P}^{2}=(M_{P^{+}}^{2}-M_{P^{0}}^{2})_{m_{u}=m_{d}}}
is the \emph{electromagnetic squared mass splitting} of the isospin multiplet $P$. One can
also consider the dimensionless \emph{relative correction to Dashen's theorem} :
\equ{\Delta_{R}D=\frac{\Delta_{\mathrm{EM}}M_{K}^{2}}{\Delta_{\mathrm{EM}}M_{\pi}^{2}}-1}
Dashen has shown in \cite{Dashen:1969p262} than $\Delta_{A}D=0$ in the $\SU(3)$ chiral
limit and that the leading corrections are $\bigo(\alpha m_{s},\alpha^{2})$.
The quantity $\Delta_{A}D$ is interesting because it is very sensitive to the
up and down quark masses.

An overview of the results for the violations to Dashen's theorem corrections is presented
in Table \ref{tab:dash}. These results are interesting for two reasons : on one hand, agreement
between different calculations is poor and on the other all these
numbers predict rather large corrections. We will investigate corrections to Dashen's theorem
using a QCD+QED analysis on BMW collaboration QCD ensembles. A first step in this
direction is to formulate electromagnetism on the lattice.

\begin{table}[h!]
\centering
\begin{tabular}{|r|c|c|l}
\cline{2-3}
\multicolumn{1}{c|}{~} & $\Delta_{A}D~(\mega\ev\squared)$ & $\Delta_{R}D$ &    \\
\cline{1-3}
\multirow{6}*{\rotatebox{90}{phenomenology}}
& $1230$                & $\mathbf{0.80}$         & \cite{Donoghue:1993p14}    \\
& $\mathbf{1300\pm400}$ & $1.02\pm 0.30$          & \cite{Bijnens:1993p12}     \\
& $\mathbf{360}$        & $0.26$                  & \cite{Baur:1996p16}        \\
& $\mathbf{1060\pm320}$ & $\mathbf{0.84\pm 0.24}$ & \cite{Bijnens:1997p11}     \\
& $\mathbf{1080}$       & $\mathbf{0.68}$         & \cite{Gao:1997p13}         \\
& $\mathbf{1070}$       & $0.85$                  & \cite{Bijnens:2007p9}      \\
\cline{1-3}
\multirow{3}*{\rotatebox{90}{lattice}}
& $526$                 & $0.39$                  & \cite{Duncan:1996p5}       \\
& $\mathbf{340\pm92}$   & $0.30\pm0.08$           & \cite{Blum:2007p8}         \\
& $\mathbf{1250\pm550}$ & N/A                     & \cite{Basak:2008p248}      \\
\cline{1-3}
\end{tabular}
\caption{Results for the violations to Dashen's theorem. Bold numbers are the results
given by the authors, the others are deduced from information given in the corresponding paper.
\label{tab:dash}}
\end{table}

%%%%%%%%%%%%%%%%%%%%%%%%%%%%%%%%%%%%%%%%%%%%%%%%%%%%%%%%%%%%%%%%%%%%%%%%%
\pagebreak
\section{Electromagnetism on the lattice}\label{sec:latem}
We impose periodic boundary condition in a finite volume, \ie we will define Maxwell's
theory on a $4$-torus $\mathbb{T}^{4}$. In this context,
the Maxwell-Gauss equations :
\equ{\sum_{\mu=0}^{3}\partial_{\mu}F_{\mu\nu}=j_{\nu}~~~\textrm{with}~~~
F_{\mu\nu}=\partial_{\mu}A_{\nu}-\partial_{\nu}A_{\mu}}
imposes global electric neutrality :
\equ{Q_{\textrm{total}}=\int_{\mathbb{T}^3}\diff^3\mathbf{x}\,j_0(x)=
    \sum_{k=1}^{3}\int_{\mathbb{T}^3}\diff^3\mathbf{x}\,\partial_{k}F_{k0}(x)=0}
One way to circumvent this global neutrality is to add a finite-volume,
constant background current density as a multiplier to the Lagrangian of the theory :
\equ{\mathscr{L}[A](c)=\frac{1}{4}\sum_{\mu,\nu=0}^{3}F_{\mu\nu}F_{\mu\nu}+
\frac{1}{V}\sum_{\mu=0}^{3}L_\mu c_\mu A_\mu\label{eq:lagr}}
where $L_{\mu}$ is the length of direction $\mu$ in $\mathbb{T}^{4}$ and
$V=L_{0}L_{1}L_{2}L_{3}$ is the 4-volume of $\mathbb{T}^{4}$. The Maxwell-Gauss equations become :
\equ{\sum_{\mu=0}^{3}\partial_{\mu}F_{\mu\nu}=j_{\nu}-\frac{1}{V}L_{\nu}c_{\nu}}
Minimizing the action with respect to the Lagrange multiplier $c$ yields a new
Euler-Lagrange equation :
\equ{\hat{A}_{\mu}(0)=\int_{\mathbb{T}^4}\diff^4x\,A_{\mu}(x)=0}
where the hat symbol denotes Fourier transform.

We formulate now electromagnetism on the lattice in its non-compact form
(following \cite{Duncan:1996p5}). This is done by using forward finite differences in the Lagrangian (\ref{eq:lagr}) :
\equ{\partial_\mu f(x)=f(x+\hat{\mu})-f(x)}
This formulation has a major advantage : 
the photon does not self-interact. The inconvenience is that 
gauge fixing becomes compulsory. We chose the following gauge conditions :
\equ{
\left\{\begin{array}{rrcl}
	\forall p_0\neq 0,& \hat{A}_0(p_0,\mathbf{0}) & = & 0\\
	\forall p,&\sum_{k=1}^{3}\widetilde{p}_k\hat{A}_k(p) & = & 0
\end{array}\right.~~~
\textrm{with}~~~\widetilde{p}_\mu=2\sin\left(\frac{p_\mu}2\right)\label{eq:gauge}}
This choice has the practical advantage that it allows to solve explicitely the gauge constraints
during field generation (\cf appendix of \cite{Blum:2007p8}).

The action $S$ associated with the above Lagrangian is a quadratic form, so generating random 
fields according to the 
law $\Diff A\,\exp(-S[A])$ to compute the path integral of the theory is simply a multi-dimensional
Gaussian random number drawing. This is done in momentum space, where the correlation matrix of
the law is quite sparse, and then the position representation of the field is recovered using 
a Fast Fourier Transform algorithm. The associated $\U(1)$ compact field (links) $U^{\mQED}_{\mu}$
is then constructed in the following way :
\equ{U^{\mQED}_{\mu}=\exp(iqA_{\mu})}
where $q$ is an electric charge.

\begin{figure}[h!]
\centering
\includegraphics[width=\plots\textwidth]{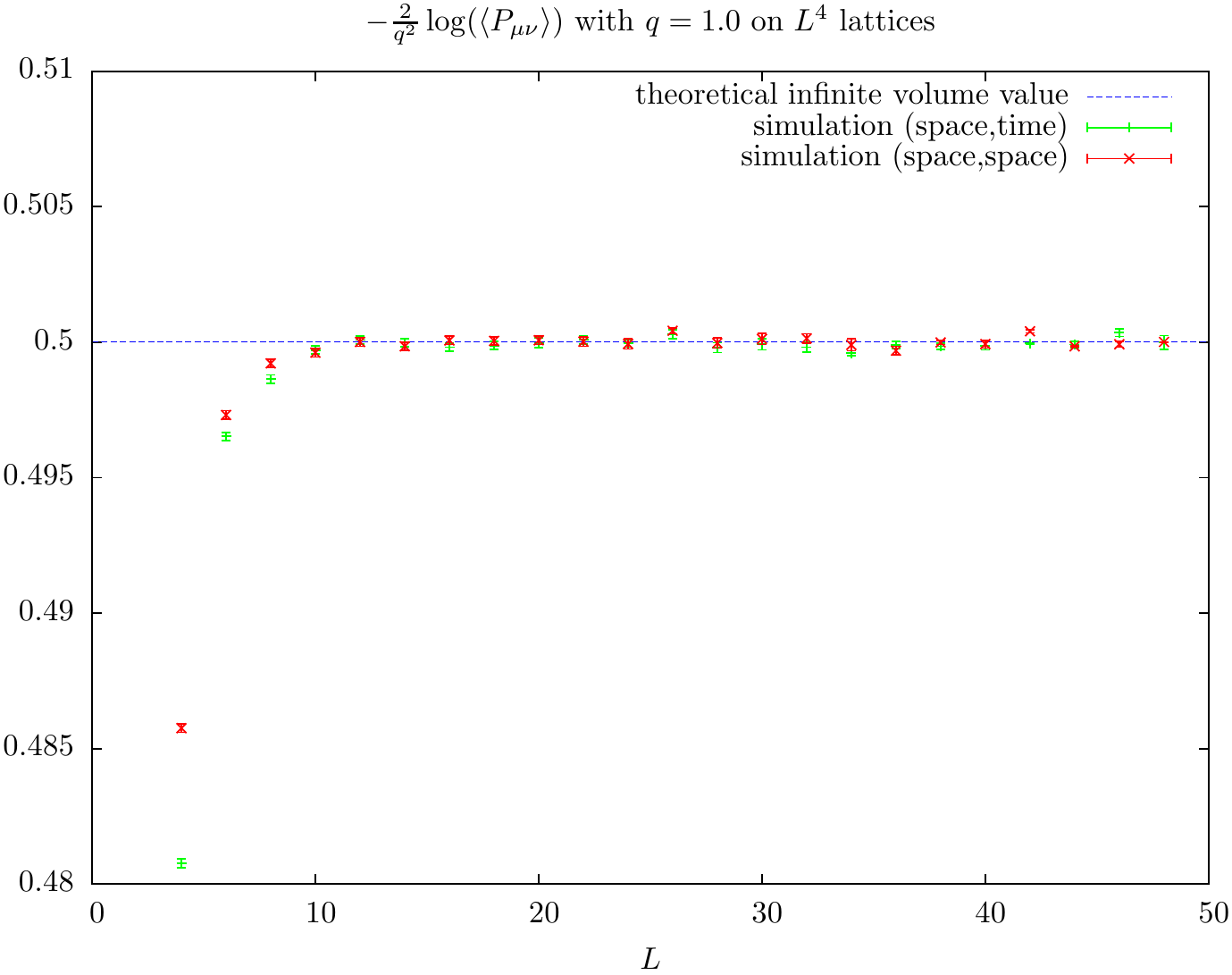}
\caption{Logarithm of the $\U(1)$ plaquette vs. lattice extent $L$. The dashed blue line represents the semi-analytical, infinite-volume prediction. The green points correspond to plaquettes aligned along the time and a space directions and the red ones to plaquettes  aligned along two perpendicular space directions.
\label{fig:u1plaq}}
\end{figure}

To check the implementation of the field generation code and investigate volume dependence,
we computed semi-analytically the expectation value of the plaquette operator $P_{\mu\nu}$ 
associated with the field $U^{\mQED}_{\mu}$ in an infinite volume
lattice to compare it with simulation values. The results are summarized in Figure \ref{fig:u1plaq}.
It appears that for lattice extents greater than $10$ the plaquette does not suffer from 
finite-volume effects and hypercubic invariance is present even with our non symmetric gauge 
choice (\ref{eq:gauge}).

%%%%%%%%%%%%%%%%%%%%%%%%%%%%%%%%%%%%%%%%%%%%%%%%%%%%%%%%%%%%%%%%%%%%%%%%%
\section{Coupling to quarks}\label{sec:coupling}
Here we describe how to perform a QCD and quenched QED analysis by re-using previously generated
$\SU(3)$ gauge configuration. We are using $N_{f}=2+1$ QCD simulations with L\"uscher-Weisz 
gauge action, 
tree level $\mathrm{O}(a)$-improved Wilson fermions and two steps of HEX smearing
(\cf \cite{Durr:2010p384,Kurth:2010} for details).

Ideally one would include the $\U(1)$ degrees of freedom directly in the Hybrid Monte-Carlo.
But doing this to reproduce the wide range of parameters already
explored in BMW collaboration ensembles is extremely expensive.
If, in a first instance, one accepts to quench QED, then there is a simple way to re-use
previously generated $\SU(3)$ fields. To compute the propagator of a quark $q$ of charge
$Q_{q}\in\left\{\frac{2}{3},-\frac{1}{3}\right\}$ proceed as follows :

\begin{enumerate}
\item generate an electromagnetic potential $A_{\mu}$ as described in Section \ref{sec:latem}
\item construct the associated $\U(1)$ field $U^{\mQED}_{\mu}=\exp(iQ_{q}eA_{\mu})$
\item using a previously generated $\SU(3)$ field $U^{\mQCD}_{\mu}$, construct the $\U(3)$ field
$U_{\mu}=U^{\mQED}_{\mu}U^{\mQCD}_{\mu}$
\item compute the quark propagator inverting the Dirac-Wilson operator using $U_{\mu}$ as the gauge
field in covariant derivatives
\end{enumerate}

\begin{figure}[h!]
\centering
\includegraphics[width=\plots\textwidth]{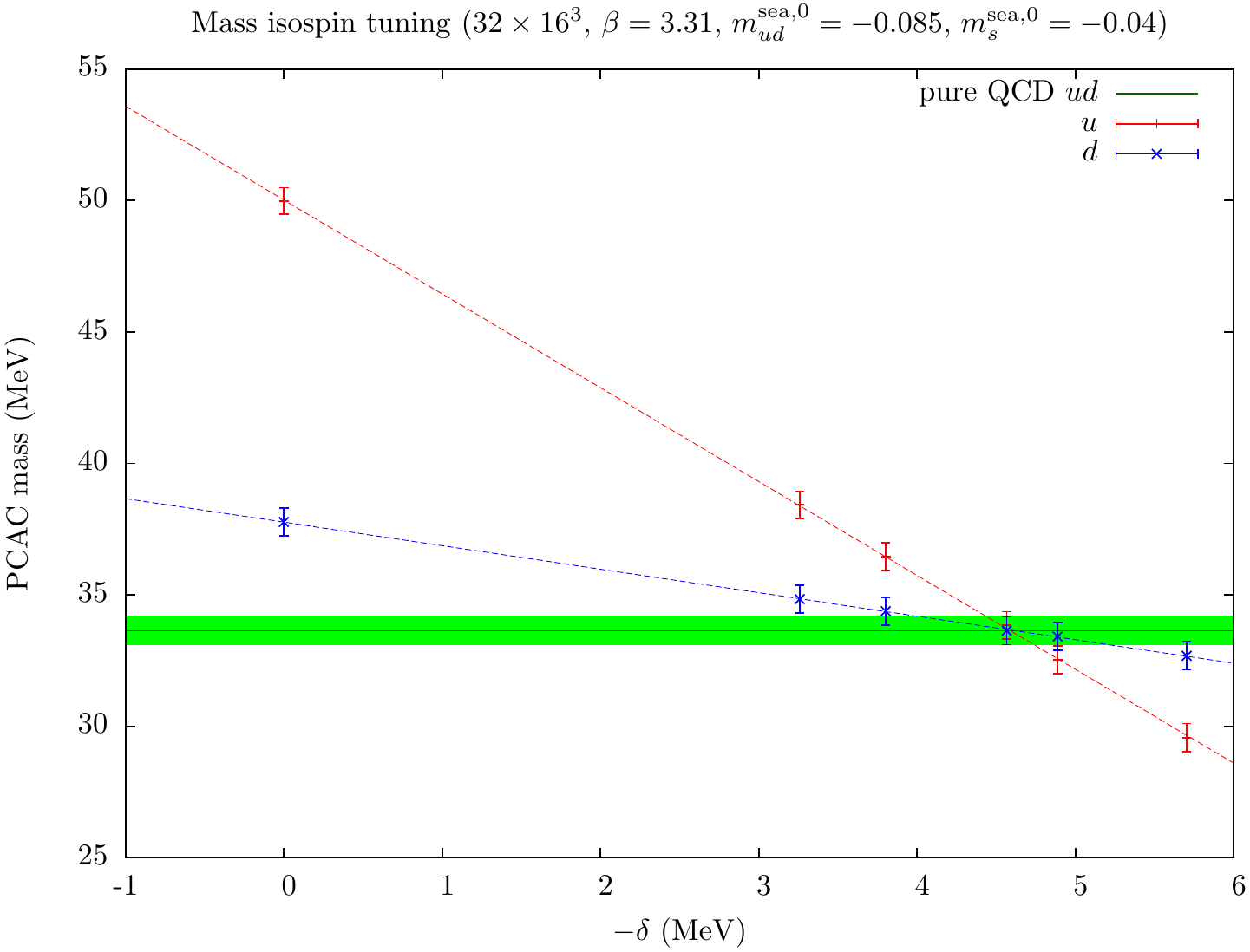}
\caption{PCAC light quark masses vs. $-\delta$. The behaviour in $\delta$ appears to be
perfectly linear. One can sees the large contribution of the $a^{-1}$ additive renormalization
 at $\delta=0$, where no substraction is applied.
\label{fig:isotun}}
\end{figure}

Because Wilson fermions break chiral symmetry explicitly, they require
an additive renormalization in the quark
masses. This additive renormalization has already been accounted for
in our pure QCD simulation. However, in the presence of an extra
$U(1)$ gauge interaction, we expect an additional additive renormalization
term, that will have to be subtracted, of order
$\bigo(Q_{q}^{2}\frac{\alpha}{a})$ where $a$ is the lattice spacing.
While suppressed by $\alpha$, this divergence may still be large
because it is enhanced by $a^{-1}$. Moreover, this correction breaks the mass isospin
symmetry of our $N_f=2+1$ simulations, since $Q_{u}\neq Q_{d}$. We
have chosen to subtract this divergence by retuning the bare $u$ and
$d$ quark masses in such a way that the renormalized masses are
equal. Thus, we substract a quantity $Q_{q}^{2}\delta$ from each
bare quark mass and tune $\delta$ to a value $\delta_{c}$ where
isospin symmetry is recovered. This isospin tuning works very well, as
shown in Figure ~\ref{fig:isotun}.

Everything is now ready to perform a QCD+QED analysis 
in the mass isospin limit to investigate
the violations to Dashen's theorem.

%%%%%%%%%%%%%%%%%%%%%%%%%%%%%%%%%%%%%%%%%%%%%%%%%%%%%%%%%%%%%%%%%%%%%%%%%
\section{Preliminary results}\label{sec:results}
Here are our first preliminary results, using the following subset of the BMW collaboration
QCD ensembles :
\begin{center}
\begin{tabular}{|c|c|c|c|c|c|c|}
\hline
$\beta$ & $m_{ud}^{0}$ & $m_{s}^{0}$ & size & $N_{\textrm{conf}}$
& $\sim M_{\pi}~(\mega\ev)$ & $M_{\pi}L$\\
\hline
$3.31$ & $-0.08500$ & $-0.04$ & $32\times 16^{3}$ & $218$ & $420$ & $4.05$\\
$3.31$ & $-0.09300$ & $-0.04$ & $48\times 24^{3}$ & $128$ & $300$ & $4.26$\\
$3.31$ & $-0.09530$ & $-0.04$ & $48\times 24^{3}$ & $210$ & $250$ & $3.61$\\
$3.31$ & $-0.09756$ & $-0.04$ & $48\times 32^{3}$ & $130$ & $200$ & $3.86$\\	
\hline
\end{tabular}
\end{center}
This subset contains a single lattice spacing estimated to be $a\simeq\unit{0.116}{\femto\meter}$
($a^{-1}\simeq\unit{1.7}{\giga\ev}$), one strange quark mass approximatively tuned to its 
physical value and four pion masses, from $\unit{420}{\mega\ev}$ down to $\unit{200}{\mega\ev}$.
Clover improvement is done only on $\SU(3)$ fields.

Time correlators of $\pi^{0}$, $\pi^{+}$, $K^{0}$ and $K^{+}$ are constructed by contracting
quarks propagators computed as described in Section \ref{sec:coupling}. The $\pi^{0}$ correlator, 
is constructed by averaging the $u\bar{u}$
and $d\bar{d}$ pseudoscalar propagators, neglecting disconnected contributions. These
contributions are subleading in $\alpha$ and $m_{u}-m_{d}$.
The mass of a pseudo-Goldstone meson is extracted fitting its time correlator to a hyperbolic
cosine. The square of this mass is then extrapolated to the physical
$\pi^{+}$ mass ($M_{\pi^{+}}^{\phi}=\unit{139.57018}{\mega\ev}$) using a Taylor expansion
around $M_{\pi^{+}}^{\phi\,2}$ (as done for \cite{Durr:2008p180}) :
\equ{M_{P}^{2}(M_{\pi^{+}}^{2})=M_{P}^{\phi\,2}\left[1+
\sum_{k=0}^{n}c_{k}(M_{\pi^{+}}^{2}-M_{\pi^{+}}^{\phi\,2})^{k}\right]} 
With this methodology, the previously described QCD ensembles and a linear extrapolation yield
the preliminary results (quoted errors are statistical only) :
\begin{eqnarray*}
M_{\pi^{0}}                     & = & \unit{134.5\pm 1.1}{\mega\ev}      \\
\Delta_{\mathrm{EM}}M_{\pi}     & = & \unit{5.1\pm 1.1}{\mega\ev}        \\
\Delta_{\mathrm{EM}}M_{\pi}^{2} & = & \unit{1380\pm 50}{\mega\ev\squared}\\
M_{K^{+}}                       & = & \unit{501.3\pm 2.0}{\mega\ev}      \\
M_{K^{0}}                       & = & \unit{499.0\pm 2.0}{\mega\ev}      \\
\Delta_{\mathrm{EM}}M_{K}       & = & \unit{2.2\pm 0.2}{\mega\ev}        \\
\Delta_{\mathrm{EM}}M_{K}^{2}   & = & \unit{2200\pm 180}{\mega\ev\squared}\\
\Delta_{A}D                     & = & \unit{830\pm180}{\mega\ev\squared}\\
\Delta_{R}D                     & = & 0.60\pm 0.14
\end{eqnarray*}

%%%%%%%%%%%%%%%%%%%%%%%%%%%%%%%%%%%%%%%%%%%%%%%%%%%%%%%%%%%%%%%%%%%%%%%%%
\section{Conclusion}
The preliminary results obtained in Section \ref{sec:results} validate our methodology 
and clearly encourage us to
perform much more detailed calculations (lower quark masses, several
lattice spacing, different volumes, \etc) to gain control over
systematic errors and to extend these calculations to other interesting hadronic observables.

%%%%%%%%%%%%%%%%%%%%%%%%%%%%%%%%%%%%%%%%%%%%%%%%%%%%%%%%%%%%%%%%%%%%%%%%%
\acknowledgments{Computations were performed using HPC resources from GENCI-[CCRT/IDRIS] 
(grant 52275) and from FZ J\"ulich, as well as clusters at Wuppertal and 
CPT. This work is supported in part by EU grants FP7/2007-2013/ERC 
$n^0$ 208740, MRTN-CT-2006-035482 (FLAVIAnet), OTKA grant AT049652 DFG grant 
FO 502/2, SFB-TR 55, U.S. Department of Energy Grant $n^0$ 
DE-FG02-05ER25681, by CNRS grants GDR $n^0$ 2921 and PICS $n^0$ 4707.}

%%%%%%%%%%%%%%%%%%%%%%%%%%%%%%%%%%%%%%%%%%%%%%%%%%%%%%%%%%%%%%%%%%%%%%%%%

\end{document}